\documentstyle[prb,aps,twocolumn]{revtex}
\def\eq{\begin{equation}}
\def\ee{\end{equation}}
\def\eqa{\begin{eqnarray}}
\def\eea{\end{eqnarray}}

\def\sxy{\sigma_{xy}} 
\def\s{\sigma}
\def\p{\partial}
\def\px{\partial_x}
\def\pt{\partial_t}

\def\w{\omega}

\def\ep{\epsilon}

\def\dt{\delta}
\begin{document}
\draft
\flushbottom
\twocolumn[
\hsize\textwidth\columnwidth\hsize\csname @twocolumnfalse\endcsname
\title{Bulk Versus Edge in the Quantum Hall Effect}
\author{Y.-C.~Kao$^*$}
\address{Theoretical Physics Group, Ernest Orlando Lawrence Berkeley National Laboratory,\\ University of California, Berkeley, 
CA 94720 \\ \rm{and}}
\author{D.-H.~Lee} 
\address{Department of Physics, University of California at Berkeley, Berkeley, CA 94720}
\date{\today}
\maketitle
\tightenlines
\widetext
\advance\leftskip by 57pt
\advance\rightskip by 57pt

\begin{abstract}
The manifestation of the bulk quantum Hall effect on edge  is the chiral anomaly.
The chiral anomaly {\it is} the underlying principle of the ``edge approach'' of quantum Hall effect.
In that approach, $\sxy$ {\it should not} be taken as the conductance derived from the space-local current-current 
correlation function of the pure one-dimensional edge problem. 
\end{abstract}
\vskip 1cm
\pacs{73.50.Jt, 05.30.-d, 74.20.-z}

]

\narrowtext
\tightenlines

The question of whether quantum Hall effect is a bulk or edge phenomena is often raised.\cite{dist} It is the purpose of this 
note to address this question. Through this note we use the unit $c=\hbar=k_B=1$.

First, we point out the connection between the Laughlin gauge argument and the edge chiral anomaly. Let us imagine sitting on a 
Hall plateau, where $\s_{xx}=0$ and $\sxy={\rm a~ quantized~value}$, and ask what is the response of such a system to Laughlin'
s flux threading.\cite{laughlin} (The geometry we consider is the ``ribbon'' used in Laughlin's original paper (see Fig.1).) In 
response to the EMF ($E_x$) induced by the changing flux, a current $\int dy J_y=\sxy\dot{\Phi}$ is induced.\cite{ss} Applying 
the Su-Schrieffer counting argument,\cite{ssch,gw} the net charge transfer is determined to be\cite{laughlin}
\eq
\dt Q=\int dtdx J_y=\sxy\int dt\dot{\Phi}=\frac{2\pi}{e}\sxy
\label{dq}
\ee
{\it Thus, the manifestation of the quantized $\sxy$, is a quantized charge transfer $\dt Q={2\pi\sxy}/{e}$.}  

>From the edge point of view, the world is chiral. Indeed, even in the absence of an applied electric field, there is a current 
flowing. Of course, from the 2D point of view, this is simply due to the combined effects of a) the slope of the spatial confini
ng potential, and b) the Hall effect. During Laughlin's gedanken experiment, a time-dependent electric field is observed along 
the edge 
\eq
E_x=\frac{\dot{\phi}}{L},
\ee
where $L$ is the circumference of the ribbon. Moreover, accompany the appearance of $E_x$,
 an influx of charge, i.e. an anomaly, occurs. The total amount of charge that flows in is given by Eq.(\ref{dq}).
 Thus a relation between $\Delta Q$ and $E_x$ can be established: 
\eq
\Delta Q=\sxy\int dxdt E_x.
\label{ana}
\ee
Eq.(\ref{ana}) is the integral form of the ``chiral anomaly''
\eq
\partial_{\mu}J^E_{\mu}=\sxy E_x.
\label{ana1}
\ee
Here $J^E_{\mu}=\pmatrix{\rho_E \cr J_E}$ is the 1+1 edge current. ($\rho_E$ and $J_E$ are the edge charge and current density 
respectively. Dimension wise, $J_E$ is the same as the total current $I$ in 2D.)
 
Thus {\it the 2D quantum Hall effect is in one-to-one correspondence with the 1D chiral anomaly.
 Moreover, the 2D Hall conductance is identical to the coefficient in front of $E_x$ in} Eq.(\ref{ana1}).
 From now on we shall refer to the latter as the ``coefficient of chiral anomaly''. The correspondence between the chiral 
anomaly in one dimension and the Chern-Simons effective action (i.e. quantum Hall effect) in two dimensions has already been 
emphasized by
Callan and Harvey.\cite{ch}

{\it The chiral anomaly is also the underlying principle of the ``edge approach'' of the quantum Hall effect}.
\cite{edge1,edge2,edge3} To see that we consider the case where the current is uniform along the edge. In that case
$\px J_E=0$ and Eq.(\ref{ana1}) becomes
\eq
\pt\rho_E=\sxy E_x.
\label{ed}
\ee
Multiplying Eq.(\ref{ed}) by the (constant) edge velocity $-v$, we obtain\cite{note10}
\eq
\pt J_E=-\sxy vE_x,
\ee
which implies
\eq
J_E(t)-J_E(0)=-\sxy\int_0^tdt'~vE_x(t').
\ee
Let us now consider the case where $E_x$ is produced by Laughlin's flux threading in time interval $(0,t)$. Moreover, let us 
assume that initially $J_E(0)=0$. At the end of the flux threading, a current
\eq
|J_E|=\sxy V
\label{edge}
\ee
is established, where $V=\int_0^t dt'~vE_x$ is the amount of work the electric field does to every unit of charge during $(0,t)$.
Another way of stating Eq.(\ref{edge}) is that if we raise the edge electro-chemical potential by $V$, a current giving by 
Eq.(\ref{edge}) will flow in the new ground state. The last statement is the building block of the edge approach used in Ref.[7-9]. 
A formula similar to Eq.(\ref{edge}), $I=\frac{e^2}{2\pi}V$, also appears in the one dimensional conduction of
 {\it free electrons} with no impurity scattering.
 In that case chiral anomaly also provides a natural intepretation of the often-confused quantized conductance.  

Next, we demonstrate that the chiral anomaly is a {\it constraint} on the edge dynamics of a quantum Hall droplet. 
First, we look at the primary QHLs ($\sxy={e^2}/{2\pi m}$), so that there is only one edge. 
We recall that the bulk effective gauge action of a QHL is
\eq
S_{eff}=\int dtd^2r [\frac{\sxy}{2}\epsilon_{abc}A_{a}\partial_{b}A_{c} + J^0_{a}A_{a}].
\label{bulk}
\ee
Throughout this paper Roman letters, e.g. $a,b,c$, are used to label the $2+1$ space-time, while Greek indices are reserved for 
the 1+1 space-time. In Eq.(\ref{bulk}) $J^0_{a}=(\bar{\rho},-v\bar{\rho},0)$ is the ground state 2+1 current,\cite{note10} and 
$\epsilon_{abc}\p_bA_c$ is the {\it perturbing} part of the external EM field.
When the Hall liquid is spatially finite, the above becomes
\eq
S_{eff}=\int dtd^2r {\cal M}(t,\vec{r})[\frac{\sxy}{2}\epsilon_{abc}A_{a}\partial_{b}A_{c} + J^0_{a}A_{a}].
\label{sup}
\ee
In the above ${\cal M}$ describes the dynamic shape of the Hall droplet, and ${\cal M}(t,\vec{r})=1$ or $0$ depending on whether 
at time $t$ the spatial point $\vec{r}$ is inside or outside the droplet.
The dynamics of ${\cal M}(t,\vec{r})$ is {\it determined} by the requirement of gauge invariance of the $S_{eff}$ in 
Eq.(\ref{sup}).\cite{wen,lw}
Trivial manipulation gives
\eq
J^0_{a}\partial_{a} {\cal M}+\frac{\sxy}{2}\epsilon_{abc}\partial_{a}{\cal M}\partial_{b}A_{c}=0.
\label{qq}
\ee
We emphasize that Eq.(\ref{qq}) is a {\it constraint} on the edge dynamics. 

Now consider the simple case where $\ep_{0ab}\p_aA_b=\ep_{1ab}\p_aA_b=0$, and a strip-like Hall droplet (Fig.1).
Let $u(x,t)$ be the normal displacement of the upper liquid boundary from the straight line, Eq.(\ref{qq}) implies
\eq
J^0_{\mu}\partial_{\mu}u=\frac{\sxy}{2}E_x,
\label{ww}
\ee
where $J^0_{\mu}\equiv (\bar{\rho},-\bar{\rho}v)$.
By identifying the chiral current (not the total edge current) as
\eq
J^C_{\mu}\equiv J^0_{\mu}u,
\label{jcc}
\ee
Eq.(\ref{ww}) becomes\cite{lw}
\eq
\partial_{\mu}J^C_{\mu}=\frac{\sxy}{2}E_x.
\label{ana9}
\ee 
The fact that the {\it chiral} current anomaly is only half of that of the {\it total} edge current is well understood.
\cite{nac,cha,bal,mae} The reason is that the total edge current is the sum of the chiral current and an additional piece. 
Indeed, if we
 solve ${\cal M}$ in terms of $\ep_{abc}\p_bA_c$ via Eq.(\ref{qq}) and substitute the answer back into Eq.(\ref{sup}),
 we obtain a gauge-invariant  effective action $S_{eff}(A_a)$.
 The total current  $J^{tot}_a=\p S_{eff}/\p A_a$ contain a bulk term and an edge one, i.e. $J^{tot}_a=J^{bulk}_a+J^{edge}_a$.
The 1+1 dimensional edge current $J^E_{\mu}(t,x)$ is obtained from the 2+1 dimensional $J^{edge}_a(t,x,y)$ via
$J^E_{\mu}(t,x)=\int dy J^{edge}_{a=\mu}(t,x,y)$.
It can easily be shown that   
\eq
\p_{\mu}J^E_{\mu}=\p_{\mu}J^C_{\mu}+\frac{\sxy}{2}\epsilon_{\mu\nu}\p_{\mu}A_{\nu}.
\label{je}
\ee
Eqs.(\ref{ana9}) and (\ref{je}) is of course equivalent to Eq.(\ref{ana1}). (In the literature $\p_{\mu}J^E_{\mu}=\sxy E_x$ is 
called the ``covariant anomaly'' while $\p_{\mu}J^C_{\mu}=\frac{\sxy}{2} E_x$ is called the ``consistent anomaly'') {\it In the 
following we shall concentrate on the consistent anomaly} (Eq.(\ref{ana9})). To obtain the covariant anomaly (i.e. the total edge 
current anomaly) we simply multiply the anomaly coefficient by 2.   

Following the approach used by Wen\cite{wen}, we now construct an edge action, so that the exact equation of motion 
reproduces Eq.(\ref{ww}). In order to get a local action, it is convenient to introduce the so-called ``chiral boson'' 
field $\phi$ so that
\eq
\bar{\rho}u=\frac{1}{2\pi}\px\phi.
\label{c}
\ee
In terms of $\phi$ the answer is (remember that $\sxy=\frac{e^2}{2\pi m}$)
\eq 
S=\int dtdx[\frac{m}{4\pi}\px\phi(\pt-v\px)\phi+\frac{e}{4\pi}\phi(\px A_t-\pt A_x)].
\label{act}
\ee
Since Eq.(\ref{act}) is quadratic in $\phi$, the saddle-point equation  given by
\eq
\px(\pt-v\px)\phi=\frac{e}{2m}(\px A_t-\pt A_x)=-\frac{e}{2m}E_x,
\label{ana2}
\ee
is exact.
Due to Eqs.(\ref{jcc}) and (\ref{c}), the above is identical to Eq.(\ref{ana9}).
Although Eq.(\ref{act}) is
derived in the spirit followed by Wen.\cite{wen}, its
gauge coupling differs from that used by Wen in important ways.
 The gauge coupling we use is dictated by the chiral anomaly (Eq.\ref{ana9}).
 We emphasize that the gauge action resulted from integrating out out $\phi$ in Eq.(\ref{act}) is
 {\it not} the edge effective action. Instead, the latter is obtained by solving ${\cal M}$ in terms of $\ep_{abc}\p_bA_c$ via 
Eq.(\ref{qq}), substituting the answer back into Eq.(\ref{sup}) and  extracting the terms that localize on the edge.\cite{lw}

The above result can be easily generalized to hierarchical QHLs. 
The effective edge action is
\eqa
S&=&\frac{1}{4\pi}\int dtdx\sum_{ij}(K_{ij}\pt\phi_i\px\phi_j-V_{ij}\px\phi_i\px\phi_j)\nonumber \\
&+&\frac{e}{4\pi}\int dtdx\sum_i t_i\phi_i(\px A_t-\pt A_x).
\label{s}
\eea
Here $\phi_i$ is the chiral boson field associated with the edge of the  ith level QHL, $K_{ij}$ is an integer-valued symmetric 
matrix, $V_{ij}$ is a positive definite matrix, and $t_i$ is the ``charge vector''\cite{wen}.
The equation of motion implied by Eq.(\ref{s}) is
\eq
\sum_j(K_{ij}\pt\px\phi_j-V_{ij}\px\px\phi_j)=-\frac{1}{2}et_iE_x.
\label{eq}
\ee
The chiral charge and current density associated with $\phi_i$ is 
\eqa
&&\rho_{c,i}=-et_i\frac{1}{2\pi}\px\phi_i\nonumber \\
&&J_{c,i}=et_i\frac{1}{2\pi}\sum_{jk}K^{-1}_{ij}V_{jk}\px\phi_k.
\label{nc}
\eea
Substituting Eq.(\ref{nc}) into Eq.(\ref{eq}) we obtain
\eq
(\frac{1}{t_i})\p_{\mu}J_{c,i\mu}=\frac{e^2}{4\pi}(K^{-1}t)_iE_x.
\ee
Thus the total chiral current anomaly is
\eq
\p_{\mu}J^C_{\mu}=\sum_i\p_{\mu}J_{c,i\mu}
=\frac{e^2}{4\pi}(t^TK^{-1}t)E_x.
\label{anoma}
\ee
Since 
\eq
\sxy=\frac{e^2}{2\pi}(t^TK^{-1}t),
\label{sxy}
\ee
Eq.(\ref{ana9}) holds.
The fact that we obtain Eq.(\ref{anoma}) is not at all surprising, since the chiral anomaly is built in as a {\it constraint} on 
the edge dynamics.

In a recent paper,\cite{kf} Kane, Fisher, and Polchinski defined a ``two terminal conductance'', from the local edge 
current-current correlation function (following that reference we shall change to the Euclidean metric below)
\eq
G=(\frac{e}{2\pi})^2|\w|\sum_{ij}t_it_j <\phi_i(-w,x=0)\phi_j(\w,x=0)>.
\label{kubo}
\ee
In the above the average on the right hand side is performed in the absence of external electric field. 
In Ref.[13] it is claimed that on a Hall plateau
\eq
G=\sxy/2.
\label{for}
\ee
Now we first show that if the QHL under consideration is primary, Eq.(\ref{for}) is indeed correct. However, for general 
hierarchical QHLs Eq.(\ref{for}) is only correct if all edge eigen modes propagate in the same direction.

By using Eqs. (\ref{act}) and (\ref{kubo}) it is simple to show that
\eq
G=\frac{e^2}{2\pi m}|\w|\int_{-\infty}^{\infty}\frac{dq}{2\pi}\frac{1}{q(-i\w+vq)}=\frac{e^2}{4\pi m},
\ee
Thus for primary QHLs the two terminal conductance defined in Eq.(\ref{kubo}) agrees with $\sxy/2$.
Is this a coincidence? To shed light on that question, we consider a hierarchical QHL.
Using Eqs.(\ref{s}) and (\ref{kubo}) it is simple to show that
\eqa
G&=&\frac{e^2}{2\pi}|\w|\int_{-\infty}^{\infty}\frac{dq}{2\pi}\frac{1}{q}\sum_{ij}t_i(-i\w K+qV)^{-1}_{ij}t_j\nonumber \\ &=&
\frac{e^2}{2\pi}\frac{|\w|}{i\w}(t^TK^{-1}Mt).
\label{e1}
\eea
In the above the matrix $M$ is given by
\eq
M\equiv\int_{-\infty}^{\infty}\frac{dq}{2\pi}(qI-i\w KV^{-1})^{-1}.
\label{e2}
\ee
Let $S$ be the linear transformation that diagonalizes $KV^{-1}$.
Thus
\eqa
&&M=S~[\int_{-\infty}^{\infty}\frac{dq}{2\pi}D]~
S^{-1},~\rm{where}\nonumber \\
&&D_{ij}=\delta_{ij}\frac{1}{q-i\w \lambda_i}.
\label{e3}
\eea
Here $ \lambda_i$ is the ith eigenvalue of $KV^{-1}$.
Now the integral can be carried out for each individual diagonal element of $D$ to yield
\eq
\int_{-\infty}^{\infty}\frac{dq}{2\pi}\frac{1}{q-i\w \lambda_i}.
=\pm \frac{i}{2}.
\label{ing}
\ee
In Eq.(\ref{ing}) the sign is plus
if $i\w \lambda_i$ lies in the upper half of the complex plane; otherwise it is minus.
To understand the physical meaning of $\lambda_i$ we
look back at Eq.(\ref{s}). In the absence of the external EM field the dispersion relation is 
\eq
\w K=qV,
\ee
or
\eq
\w KV^{-1}=q I,~~I={\rm identity~matrix}.
\ee
If $K$ and $V$ are $N\times N$ matrices, there are $N$ solutions
\eq
\w=\lambda_i^{-1} q~~i=1,...,N.
\ee
Thus ${ \lambda_i}$ is the inverse velocity of the ith eigen mode, consequently it should be real. 
Therefore 
\eq
\int_{-\infty}^{\infty}\frac{dq}{2\pi}\frac{1}{q-i\w \lambda_i}.
=\frac{i}{2}\frac{\w}{|\w|}\frac{\lambda_i}{|\lambda_i|},
\label{ing1}
\ee
and 
\eq
[\int_{-\infty}^{\infty}\frac{dq}{2\pi}D]_{ij}=\delta_{ij}\frac{i}{2}\frac{\w}{|\w|}\frac{\lambda_i}{|\lambda_i|}.
\ee
Substituting the above result into Eqs. (\ref{e1})-(\ref{e3}) we obtain
\eqa
&&G=\frac{e^2}{4\pi}(t^TK^{-1}S\Lambda S^{-1}t),~~{\rm where}\nonumber \\
&&\Lambda_{ij}=\delta_{ij}\frac{\lambda_i}{|\lambda_i|}.
\eea
A great simplification occurs if all $\lambda_i$ are positive.
In that case $\Lambda=I$, and
\eq
G=\frac{e^2}{4\pi}(t^TK^{-1}t)=\frac{1}{2}\sxy.
\ee
However, in general, when $\lambda_i$ of both sign exists, $G\ne\frac{1}{2}\sxy$. For example as shown In Ref.[13], for 
the $\nu=2/3$ QHL, 
\eq
K=\pmatrix{1 & 0\cr 0 & -3},~~~V=\pmatrix {v_1 & v_{12} \cr v_{12} & v_2},~~~t=\pmatrix{1 \cr 1}
\ee
one can show that $G=\frac{\Delta}{3\pi}$, where
$\Delta=\frac{2-\sqrt{3}c}{\sqrt{1-c^2}}$ with $c=\frac{2v_{12}}{\sqrt{3}(v_1+v_2)}$.
However, in Ref.[13] this result was taken as the indication that another mechanism (edge impurity scattering) has to be invoked 
to yield a quantized $\sxy$.
Our message is that it is the coefficient of chiral current anomaly (Eq.(\ref{ana9})) instead of $G$ 
(Eq.(\ref{kubo}) that should be identified with $\frac{1}{2}\sxy$. This point has already been emphasized by 
Haldane\cite{haldane}, and by Nagaosa and Kohmoto\cite{kohmoto} 

Thus we find that {\it the bulk and edge pictures of quantum Hall effect are totally consistent}. 
The bulk quantum Hall effect corresponds to the edge chiral anomaly. The quantization of the bulk 
Hall conductance is manifested as the quantization of the chiral anomaly coefficient. 
Finally we ask ``under what condition is the edge theory used above the correct low energy description?''
 Since the edge theory is a direct consequence of the bulk quantum Hall effect (Eq.(\ref{bulk})),
 the question reduces to ``to what extent is Eq.(\ref{bulk}) the correct bulk effective action?''  
One way to view the stability of the bulk quantum Hall effect is through the boson Chern-Simons theory.\cite{zhk,lz,lf} In that 
theory, the quantum Hall effect is explained in terms of the superconductivity of composite bosons. For example, the composite 
boson for the $\nu=1$ plateau is made up of an electron bound to a fictitious magnetic flux quantum.
 When the composite boson condense, the $\nu=1$ quantum Hall effect is exhibited.
 However, when the vortices of the composite boson condense, the system becomes insulating.\cite{klz} Wen's bulk effective gauge 
theory is the dual form of the boson Chern-Simons theory upon abandoning the vortices in infrared limit.\cite{lf}. Thus, Wen's 
action will continue be the low energy effective theory, as long as 
the vortices of the Chern-Simons boson do not condense.
 Under that condition, the effective edge theory discussed above remains valid, and the chiral anomaly coefficient remains 
unchanged.\cite{note20} 
Of course, the real tough question is whether a particular condition will cause the vortices of composite boson to condense. This 
is a localization issue which is beyond the scope of this paper.
\vspace{0.2in}
\\
Acknowledgement: We thank Dr. S.A. Kivelson for helpful discussions. YCK is supported in part by the National Science Council of 
Taiwan under grant number NSC85-2112-M-002-009.
DHL is support in part by the LDRD program under DOE contract number DE-AC03-76SF00098.
\vskip 0.2cm
\noindent$^*$ Permanent address: Department of Physics, National Taiwan University, Taipei, Taiwan, R.O.C.

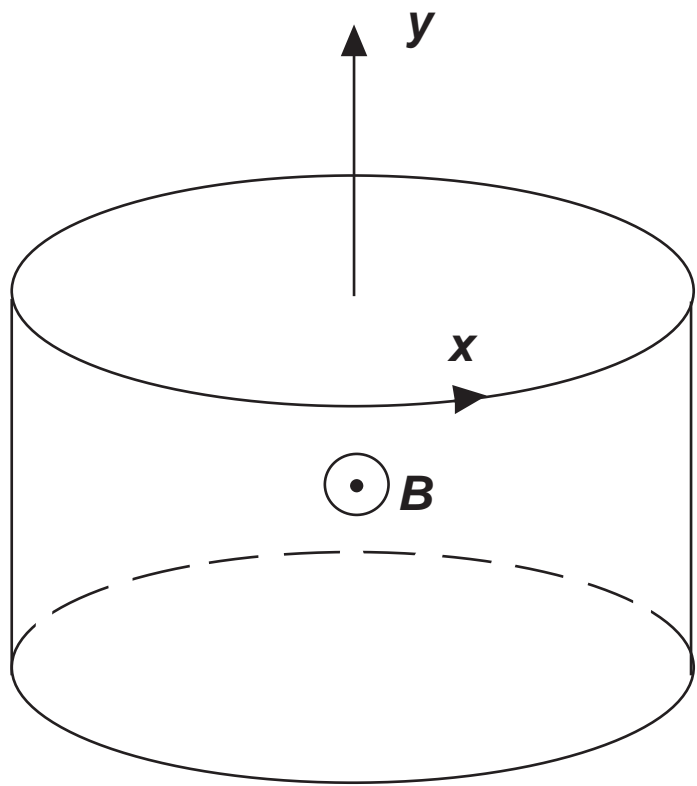
\begin{figure}
\caption{The upper edge (situated at $y=0$) of the ribbon is under consideration. The edge velocity is along $-\hat{x}$, the 
induced EMF is along $\hat{x}$, and $\dot{\Phi}$ is along $-\hat{y}$.}
\label{fig1}
\end{figure}

\begin{references}
\bibitem{dist} We emphasize that in this paper we do not address the issue of the distribution of Hall current. For the latter 
non-linear effects, which we have ignored, are essential.
\bibitem{ss} We emphasize that here we have used the basic fact of quantum Hall effect, i.e., $\s_{xx}=0$ and $\sxy\ne 0$.
\bibitem{note10} The minus sign in the second equation is due to the fact that under the geometry we defined earlier, the edge 
current flows in the $-\hat{x}$ direction.
\bibitem{note20} The question on the finite-size correction
of $\sxy$ is a more delicate issue which we do not discuss in this paper.
\end{references}
\end{document}